# Classification and Performance of AQM-Based Schemes for Congestion Avoidance

K.Chitra
Lecturer, Dept. of Computer Science
D.J.Academy for Managerial Excellence
Coimbatore, Tamil Nadu, India – 641 032
chitrakandaswamy@yahoo.com

Dr. G.Padamavathi
Professor & Head, Dept. of Computer Science
Avinashilingam University for Women,
Coimbatore, Tamil Nadu, India – 641 043

*Abstract*— Internet faces the problem of congestion due to its increased use. AQM algorithm is a solution to the problem of congestion control in the Internet. There are various existing algorithms that have evolved over the past few years to solve the problem of congestion in IP networks. Congested link causes many problems such as large delay, underutilization of the link and packet drops in burst. There are various existing algorithms that have evolved over the past few years to solve the problem of congestion in IP networks. In this paper, study of these existing algorithms is done. This paper discusses algorithms based on various congestion-metrics and classifies them based on certain factors. This helps in identifying the algorithms that regulate the congestion more effectively

I. INTRODUCTION

Today's world is dominated by Internet which results in high Internet traffic. Firstly, Internet is no longer a small, closely interleaved user community but expanded to a very large community network resulting in increased Internet traffic. Secondly, the increased use of multimedia applications also results in bursty flows in the Internet. So there is a requirement of regulating bursty flows in the very large network, the Internet. To regulate these bursty flows, resource allocation must be done efficiently. The resource allocation can be taken care by either end sources or by the network itself.

In this paper the strategies or schemes discussed moves the burden of the resource utilization or allocation to the network itself rather than the end sources. Resource utilisation must be intelligently done inside the network for efficient flow in the internet. In a network each router uses finite buffer or queue for the packets to be stored and transmitted. As a result network gets congested in case of heavy traffic and due to unresponsive and non TCP-compatible flows the danger of congestion and collapses the network. Now a days real-time Internet application like video conferencing floods the Internet routers with data that requires efficient buffer management.

Queue management in routers plays an important role in taking care of congestion. Two approaches are adopted to solve this problem. First one is Congestion Avoidance preventive technique, which comes into play before network is congested by overloading. Second is Congestion Control, which comes into play after the congestion at a network has occurred and the network is overloaded. A congestion avoidance scheme is a proactive one that maintains the network in a state of low delay and high throughput by keeping the average queue size low to accommodate bursty traffic and transient congestion. It makes TCP responsive to congestion, as TCP will back off its transmission rate when it detects packet loss. However the second one is a reactive scheme that reacts after the congestion occurs.

The two main objectives of queue management is high link utilisation with low packet loss and low packet queuing delay. These objectives conflict with each other. A small buffer can guarantee low queuing delay but it suffers from high packet loss and low link utilisation. Hence the problem arises of how to manage queue in a router. Queue management is strongly associated with packet drop. So the question that arises is when to drop a packet and which one to drop. The traditional scheme used for queue management is the passive queue management that is a congestion control approach. FIFO drop-tail [20] is one of the traditional schemes for passive queue management. According to passive queue management, packets are dropped only when the buffer is full. This scheme results in high packet loss and long queuing delay. It also introduces lock out problem and global synchronization. The congestion control approach is not suited to interactive network applications such as voice-video session and web transfers requiring low end-to-end delay and jitter because the drop-tail queue are always full or close to full for long periods of time and packets are continuously dropped when the queue reaches its maximum length. So delay will be large which will make interactive applications unsustainable. Second major disadvantage of drop tail is the global synchronization problem, which arises because the full queue length is unable to absorb bursty packet arrivals and thus many of them are dropped resulting in global synchronization. Thus, global synchronization causes all the sources to slow down at the same time resulting in long periods of low link utilization. Moreover, another main reason for global synchronization is lockout behavior of drop tail where the queue is monopolized by some flows and other connections may not easily use the queue.

To remove such problems, Active Queue Management (AQM) has been introduced in recent years that is a congestion avoidance preventive approach. The first AQM algorithm RED detects congestion by observing the queue state. In RED [2] [10] packet drop probability is linearly proportional to queue length. The AQM algorithm RED drops packets before a queue becomes full. This reduces the number of packets dropped. RED and its variant uses queue length as a congestion indicator that results in certain drawbacks. In order to overcome the difficulty of relying only on queue length to identify the level of congestion various other AQMs are introduced with different congestion indicators.

To overcome these problems with RED, REM [1] was proposed. This AQM scheme attempts to make user input rates equal to the network capacity. In case of high congestion, sources are indicated to reduce their rates. In contrast to RED, REM decouples congestion measure from performance measure which stabilizes the queue around its target independent of traffic load leading to high utilisation and low delay. AQM schemes like GREEN [8], AVQ [15] also depend on arrival rate to control the congestion in the router. AVQ uses only the traffic input rate for the measure of congestion. This provides early feedback of congestion. It provides a better control than the number of other well known AQM schemes.

Another AQM scheme BLUE [6] does not use queue length as a congestion metrics. BLUE uses packet loss and link utilization as a congestion indicator. BLUE improves RED's performance in all the aspects. It is extremely simple and provides a significant performance improvement over the RED queue. This AQM maximizes the link utilisation but suffers from large queuing delays. In LRED [24] packet loss ratio is used to design more adaptive and robust AQM. It uses the instantaneous queue length and packet loss ratio to calculate the packet drop probability. In section II, a comprehensive survey of all possible AQM schemes is presented. The main idea is to track the basic schemes that exist and classify them based on congestion metric and flow information. This section exhibits a classification of AQM schemes with the study of each AQM. In section III the various algorithms are compared, analyzed and discussed to identify the better AQM algorithms in terms of performance metrics. The section IV summarizes the previous section.

## II. BACKGROUND

In recent years, research activities have come out with various congestion avoidance mechanisms in Internet to completely avoid congestion or to improve Internet traffic. Each of these mechanisms is inefficient in certain circumstances especially in heavy traffic network that research bas become a continuous process in identifying the best Active Queue Management algorithm. Congestion in routers results in high packet loss leading to high cost that is reduced by the various existing AQM schemes.

The existing schemes use various factors or metrics to detect congestion. These factors are used to estimate congestion in the queue based on which various AQM algorithms are proposed in the past few years. The schemes are based on congestion metrics like Queue-length, Load, both Queue and Load, others like Loss rate. Further some of these schemes also use flow information along with various congestion metrics to analyze and control the congestion in routers more accurately. Considering these factors AQM schemes can be categorized based on congestion metrics without flow information and with flow information as shown in Fig 1.

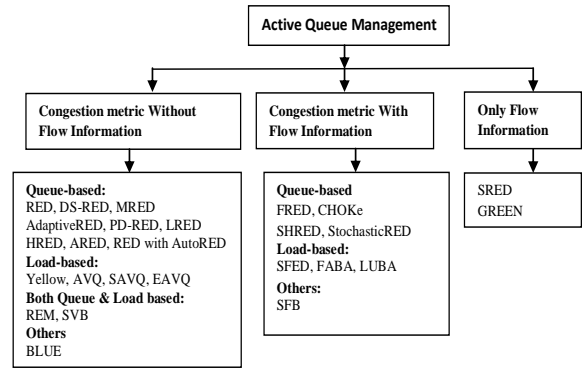

Figure 1 Classification of AQM Schemes

### A. Congestion metric without Flow Information

It is the first category of classification that considers only the congestion metric and not the flow information. However, based on the congestion metric further the AQMs can be classified. AQMs use a variety of congestion metrics like Queue length, load and link utilization to sense the congestion in routers.

#### 1) Queue-based AQM

*a) RED:* The first well known AQM scheme proposed is RED. It is one of the popular algorithms. It tries to avoid problems like global synchronization, lock-out, bursty drops and queuing delay that exists in the traditional passive queue management i.e Droptail scheme.

The algorithm in Fig. 2 detects congestion by computing the average queue size $Q_{ave}$. To calculate average queue size, low pass filter is used which is an exponential weighted moving average (EWMA). The average queue is then compared with two thresholds: a minimum threshold $min_{th}$ and a maximum threshold $max_{th}$. If the average queue size is between minimum and maximum threshold, the packet is dropped with a probability. If it exceeds maximum threshold, then the incoming packets are dropped. Packet drop probability is linear function of queue length. So the dropping probability depends on various parameters like $min_{th}$, $max_{th}$, $Q_{ave}$ and $w_q$. These parameters must be tuned well for the RED to perform better. However, it faces weaknesses such as accurate parameter configuration and tuning. This becomes a major disadvantage for the RED algorithm. Though RED avoids global synchronization but fails when load changes dramatically. Queue length gives minimum information regarding the severity of congestion. RED does not consider the packet arrivals from the various sources, which is also a very important measure for the congestion indication.

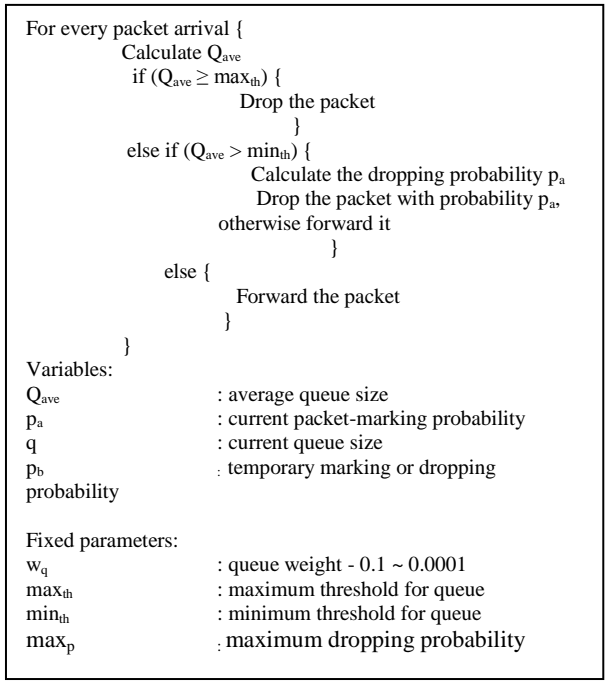

```
For every packet arrival {
        Calculate Q_ave
        if (Q_ave ≥ max_th) {
                Drop the packet
        }
        else if (Q_ave > min_th) {
                Calculate the dropping probability p_a
                Drop the packet with probability p_a,
            otherwise forward it
        }
        else {
                Forward the packet
        }
}
Variables:
Q_ave                : average queue size
p_a                  : current packet-marking probability
q                    : current queue size
p_b                  : temporary marking or dropping
probability

Fixed parameters:
w_q                  : queue weight - 0.1 ~ 0.0001
max_th               : maximum threshold for queue
min_th               : minimum threshold for queue
max_p                : maximum dropping probability
```

Figure 2 Pseudocode of the RED algorithm

Since RED considers only the queue length and not interpacket arrivals, the congestion remains as an inherent problem. In case of number of users increasing, the performance of the RED queue degrades.

According to queuing theory, it is only when packet inter-arrivals have a Poisson distribution that queue length directly relate to the number of active sources and thus indicating the true level of congestion. However in network gateways packet inter-arrival times are decided non-Poisson which clearly does not indicate the severity of congestion.

Packet loss and utilization at the link varies with regard to the network load variation as RED is sensitive to parameter configuration. In case of accurate tuning of parameter $w_q$, high utilization and low packet drop at the link can be achieved. In case of poor $min_{th}$, poor utilization at the link exists and poor $max_{th}$ value results in large packet drop

*b) DS-RED:* RED uses a single linear drop function to calculate the drop probability of a packet and uses four parameters and average queue to regulate its performance. RED suffers unfairness and low throughput. DS-RED [27] uses two-segment drop function which provides much more flexible drop-operation than RED. However, DSRED is similar to RED in some aspects. Both of them use linear drop functions to give smoothly increasing drop action based on average queue length. Next they calculate the average queue length using the same definition. The two segment drop function of DSRED uses the average queue length which is related to long term congestion level. As the congestion increases, drop will increase with higher rate instead of constant rate. As a result, congestion will be relieved and throughput will increase. This results in a low packet drop probability at a low congestion level and gives early warning for long term congestion. DSRED showed a better packet drop performance resulting in higher normalized throughput than RED in both the heavy load and low load. It results in lower average queuing delay and queue size than RED.

*c) MRED:* To overcome problems faced in RED, MRED [14] computes the packet drop probability based on a heuristic method rather than the simple method used in RED. In this scheme the average queue size is estimated using a simple EWMA in the forward or backward path. The packet drop probability is calculated to determine how frequently the router drops packets at the current level of congestion. In MRED the packet drop probability is computed step form by using packet loss and link utilization history. MRED is able to improve fairness, throughput and delay compared to RED.

*d) AdaptiveRED:* The AdaptiveRED as proposed in [9] uses the congestion indicator as the queue length. It overcomes the drawback that exists in RED that requires constant tuning of parameters depending on the traffic conditions in the network. AdaptiveRED removes this dependency by auto-tuning the parameters $w_q$ and $max_p$. The value of these parameters varies based on the network condition and keeps the average queue size within a target range halfway between the threshold $min_{th}$ and $max_{th}$. The general design of this algorithm is $w_q$ is automatically set based on the network capacity and the $max_p$ is adapted based on the measured queue length. This algorithm maintains the average queue size within a predetermined range by adapting slowly and infrequently using the Additve Increase Mulitplicative Decrease policy. The main problem of RED is parameter tuning to adapt to suit the network condition. This is automatically done in ARED by adapting $w_q$ and $max_p$ for varying network conditions to improve the performance of network. It regulates the queue utilization and packet loss rate by influencing the value of the $w_q$ and $max_p$. This gives a better result than RED with increased throughput, reduced packet loss and a predictable queuing delay.

*e) PD-RED:* PD-RED [23] was introduced to improve the performance over the Adaptive RED scheme. This scheme is based on the proportional derivative (PD) control principle. It includes control theory and adapts the maximal drop rate parameter to RED called $max_p$ to stabilise the queue length. In this scheme, AQM is considered as a typical control system. PD-RED algorithm is composed of two parts a new PD controller and the original RED AQM. The variation of queue length and the drop probability is smaller in PD-RED compared to Adaptive RED. PD-RED showed better performance in terms of mean queue length and standard deviation of the queue length.

*f) LRED:* The AQM scheme Loss Ratio based RED, measures the latest packet loss ratio, and uses it as a complement to queue length in order to dynamically adjust packet drop probability. So in this scheme packet loss ratio is a clear indication of severe congestion occurance. Queue-length is also used in small time-scale to make the scheme more responsive in regulating the length to an expected value.

LRED tries to decouple the response time and packet drop probability, there making its response time almost independent of network status.

*g) HRED* [11]: In RED, the drop probability curve is linear to the change of the average queue size. But in this paper, the drop probability curve is a hyperbola curve. As a result this algorithm regulates the queue size close to the reference queue value. This makes the algorithm no longer sensitive to the level of network load, low dependency on the parameter settings. It also achieves higher network utilization. Since HRED is insensitive to the network load and queue size does not vary much with the level of congestion, the queueing delay is less unpredictable. It rapidly reaches and keeps around its reference queue length, irrespective of the increase or decrease in queue length. Hyperbola RED tries to provide the highest network utilization because it strives to maintain a larger queue size.

*h) ARED:* This is an adaptive RED controller designed to offer better performance, adopts a self-tuning structure to try to keep the average queue length of RED gateway around the target value. The maximum drop probability is adaptively adjusted using the gradient descent method based on discrete deterministic mathematical model of TCP/RED. When the queue length in the router buffer exceeds the minimum threshold of ARED [25], the self-tuning function is used to adjust the maximum drop probability. It behaves well under light, heavy as well as changing network load conditions. When the queue size is stabilized around the optimal value, a good tradeoff between throughput and delay is achieved.

*i) AutoRED:* The AutoRed feature takes care of the traffic properties, congestion characteristics and the buffer size. In AutoRed [22], calculating the average queue size using EWMA model is modified and redefined. Therefore $w_{q,t}$ is a combination of the three main network characteristics such as traffic properties, congestion characteristics and the queue normalization. In the above technique, the $w_{q,t}$ is written as a product of the three network characteristics. The AutoRed with RED performs better than the RED scheme. This model reduces the queue oscillations appropriately in the RED-based algorithms. The AutoRed uses the strength and effect of both the burstiness and the transient congestion.

*2) Load-based AQM*

*a) AVQ:* The virtual queue is updated, when a packet arrives at the real queue to indicate the new arrival of the packet. As in Fig 3 when the virtual queue or buffer overflows, the packets are marked / dropped. The virtual capacity of the link is modified such that total flow entering each link achieves a desired utilization of the link.

This is done by aggressive marking when the link utilization exceeds the desired utilization and less aggressive when the link utilization is below the desired utilization. As a result this provides early feedback than the RED.

YELLOW: In this scheme [17], routers periodically monitor their load on each link and determine a load factor, the available capacity and the queue length. This helps in

```
At each packet arrival epoch do
        VQ = max(VQ - Ĉ(t - s), 0)       /*Update Virtual Queue
                                            Size */
    If VQ + b > B
                    Mark or drop packet in the real queue
    else
                    VQ = VQ + b
                        /* Update Virtual Queue Size */
    endif
    Ĉ = max(min(Ĉ + α * γ * C * (t - s),C) – α* b,0)
    s = t
Constant
C = Capacity of a link
B = buffer size
b = number of bytes in current packet
α = smoothing parameter.
γ = desired utilization of the link
Other
Ĉ = Virtual queue capacity
t = Current time
s = arrival time of previous packet
VQ = Number of bytes currently in the virtual queue
```

Figure 3. Pseudocode of AVQ

identifying the incipient congestion in advance and calculates the packet marking probability. Yellow improves the robust performance with respect to round-trip propagation delay by introducing the early queue controlling function. So Yellow uses the load factor (link utilization) as a main merit to manage congestion. To improve congestion control performance, a queue control function (QCF) is introduced as a secondary merit. The sufficient condition for globally asymptotic stability is presented based on Lyapunov theory. Furthermore, the principle for parameter settings is given based on the bounded stable conditions.

*b) SAVQ:* It is observed that the desired utilization parameter γ in AVQ algorithm has an influence on the dynamics of queue and link utilization. It is difficult to achieve a fast system response and high link utilization simultaneously using a constant value γ. An adaptive setting method for γ is proposed according to the instantaneous queue size and the given reference queue value. This new algorithm, called stabilized AVQ (SAVQ) [18], stabilizes the dynamics of queue maintaining high link utilization.

*c) EAVQ:* It is a rate based stable enhanced adaptive virtual queue proposed in paper [26]. Arrival rate at the network link is maintained as a principal measure of congestion. A subordinate measure is used as the desired link utilization to solve the problem such as hardness of parameter setting, poor ability of anti-disturbance and a little link capacity low. The EVAQ proved the transit performance of the system and assured the entire utilization of link capacity. Based on linearization, the local stability conditions of the TCP/EAVQ system were presented. The simulation results show the excellent performances of EAVQ such as the higher utilization, the lower link loss rate, the more stable queue length, and the faster system dynamic response than AVQ.

*Queue and Load-based AQM*

*d) REM:* As discussed Random Exponential Marking (REM) achieves high utilization with negligible loss or queuing delay even as the load increases. This scheme stabilizes both the input rate around link capacity and the queue around a small target independent of the number of users sharing the link. It uses a congestion measure price to determine the marking probability. The congestion measure price is updated based on the rate mismatch and queue mismatch as in Fig. 4.

---

$p_l(k + 1) = [p_l(k) + \gamma(\alpha l(b_l(k) - b_l^*) + x_l(k) - c_l(k))]^+$

Constants
$\gamma > 0$
$\alpha l > 0$
$b_l^*$     : target queue length
$b_l(k)$   : aggregate buffer occupancy
$c_l(k)$   : available bandwidth

---

Figure 4. Calculation of congestion measure price

When the number of users in the network increases, the queue mismatch and rate mismatch increases increasing the price value. Increase in price value results in increased marking probability. This in turn reduces the source rate of the user input. When the source rates are too small, the mismatch is negative, decreasing the price and marking probability value that increases the source rate. The price adjustment rule tries to regulate user rates with network capacity and controls queue length around a target value. RED tries to couple the congestion measure and the performance measure, but REM decouples the congestion measure and the performance measure showing a better performance than the earlier scheme.

*e) SVB:* The SVB [5] scheme uses the packet arrival rate and queue length information to detect congestion in an Internet router. As AVQ, it maintains a virtual queue and responds to the traffic dynamically. A new packet arrival is reflected in the virtual queue considering both the queue length and the arrival rate. The most striking feature of the proposed scheme is its robustness to workload fluctuations in maintaining a stable queue for different workload mixes (short and long flows) and parameter settings. The service rate of the virtual queue is fixed as the link capacity of the real queue and adapts the limit of the virtual buffer to the packet arrival rate. The incoming packets are marked with a probability calculated based on both the current virtual buffer limit and the queue occupancy. The simulations results have shown that it provides lower loss rate, good stability and throughput in dynamic workloads than the other AQM schemes like RED, REM and AVQ.

*3) Others Congestion metrics (Loss event, Link history, link utilization)*

*a) BLUE:* The BLUE algorithm resolves some of the problems of RED by employing two factors: packet loss from queue congestion and link utilization. So BLUE performs queue management based on packet loss and link utilization as shown in Fig. 5. It maintains a single probability $p_m$ to mark or drop packets. If the buffer overflows, BLUE increases $p_m$ to increase the congestion notification and is decreased to reduce the congestion notification rate in case of buffer emptiness.

This scheme uses link history to control the congestion. The parameters of BLUE are $\delta_1$, $\delta_2$ and freeze time. The freeze time determines the minimum time period between two consecutive updates of $p_m$.

BLUE maintains minimum packet loss rates and marking probability over varying queue size and number of connections compared to RED. In case of large queue, RED has continuous packet loss followed by lower load that leads to reduced link utilization.

---

Upon Packet loss (or Qlen > L) event:
if ( ( now − last_update) > freeze_time )
        $p_m := p_m + \delta_1$
        last_update := now
Upon link idle event:
if ( ( now − last_update) > freeze_time)
        $p_m := p_m - \delta_2$
        last_update := now
Constant:
$\delta_1, \delta_2$
freeze_time     : minimum time period between two consecutive updates of $p_m$

---

Figure 5 Pseudo code of BLUE algorithm

In BLUE, the queue length is stable compared to RED, which has a large varying queue length. This ensures that the marking probability of BLUE converges to a value that results in reduced packet loss and high link utilization.

*B. Congestion metric With Flow Information*

AQMs also belong to this category using both congestion metric and the flow information to detect congestion in routers. AQMs that used only congestion metric and not flow information faced the problem of unfairness in handling the different types of traffic. While considering the congestion metric they can be further classified as Queue-based or load based and others.

*1) Queue-based*

*a) FRED:* This is based on instantaneous queue occupancy of a given flow. It removes the unfairness effects found in RED. FRED [16] generates selective feedback to a filtered set of connection having a large no. of packets queue rather than choosing connections randomly to drop packets proportionally. It provides better protection than RED for adaptive flows and isolating non-adaptive greedy flows.

*b) CHOKe:* CHOKe (CHOose and Keep for responsive flows, and CHOose and Kill for unresponsive flows) [21] algorithm penalizes misbehaving flows by dropping more of their packets. So CHOKe tries to bring fairness for the flows that pass through a congested router.

CHOKe in Fig. 6 calculates the average occupancy of the buffer like as in RED using EWMA. If average queue is greater than $min_{th}$, the flowid of each arriving packet and a randomly selected packet called drop candidate packet is compared. If the packets are of the same flow then the drop both the packets. Otherwise if average queue is greater than $max_{th}$, then drop the new packet else place the packet in the buffer and admit the new packet with a probability p

```
Calculate Q_ave
if (Q_ave ≤ min_th) {
        Admit new packet
                }
else {
      Draw a drop candidate packet at random from buffer.
      If flowid of arriving packet and drop candidate packet is
      same
                Drop both packets
      else
                if (Q_ave ≤ max_th)
                        Admit the packet with probability p
                else
                        Drop the new packet.
     }
```

Figure 6 Pseudo code of CHOKe algorithm

*c) SHRED:* Short-lived flow friendly RED (SHRED) [3], an AQM mechanism improved response time for short lived Web traffic. It uses a cwnd hint from a TCP source to compute the cwnd ratio of an arriving packet to the cwnd average and reduces the probability of dropping packets during the sensitive period when a flow's cwnd is small. Sources mark each packet with its current window size, allowing SHRED to drop packets from flows with TCP windows with a lower probability. Small TCP window sizes can significantly affect short-lived flows. A small TCP window results in a lower transmission rate and short-lived flows are more sensitive to packet drops. SHRED provides improvement in web response time and is web traffic performance improvements are achieved without negatively impacting long-lived FTP traffic.

*d) Stochastic RED:* To handle the tremendous growth of unresponsive traffic internet, Stochastic RED [4] was introduced. Basically, StoRED tunes the packet drop probability of RED for all the flows by taking into consideration the bandwidth share obtained by the flows. The dropping probability is adjusted such that the packets of the flow with high transmission rate are more likely to be dropped than flows with lower rate. This algorithm distinguishes individual flows without requiring per-flow state information at the routers. It is called stochastic because it does not really distinguish the flows accurately. The arriving traffic is divided by the router into a limited number of counting bins using a hashing algorithm. On the arrival of each packet at the queue, a hash function is used to assign the packet to one of the bins based on the flow information. It dispatches the packets of the different flows to the set of bins. With a given hash function, packet of the same flow are mapped to the same bin. Therefore when the flow is unresponsive, the bin load increases dramatically.

Stochastic RED estimates the bin loads and uses these loads to penalize flows that map to each bin according to the load of the associated bin. Thus unresponsive flows experience a large packet drop probability. The StoRED is effective in disciplining misbehaving flows, making unresponsive flows TCP friendly and improving the response time of Web transfer without degrading the link utilisation.

*2) Load based*

*a) SFED:* SFED [13] is rate control based AQM discipline which is coupled with any scheduling discipline. It maintains a token bucket for every flow or aggregate flows. The token filling rates in proportion to the permitted bandwidths. When a packet is enqueued, tokens are removed from the corresponding bucket. The decision to enqueue or drop a packet of any flow depends on the occupancy of its bucket at that time. A token bucket serves as a control on the bandwidth consumed by a flow. SFED ensures early detection and congestion notification the adaptive source. The token bucket also keeps record of the bandwidth used by its corresponding flow in the recent past.

*b) FABA:* The AQM scheme fair bandwidth allocation [12] provides fairness amongst competing flows even in the presence of the non-adaptive flows. It is a rate control based AQM algorithm. It offers congestion avoidance by early detection and notification with low implementation complexity. It maintains per active-flow state with scalable implementation. It performs better than RED and CHOKe. In case of buffer sizes constrained, it performs significantly better than FRED. It gives high values of fairness for diverse applications such as FTP, Telnet and HTTP. Performance is superior even for a large number of connections passing though the routers. It is a scalable algorithm.

*c) LUBA:* LUBA [19] is link utilization based AQM algorithm. In this algorithm malicious flows are identified which causes congestion at the router, and assigns them drop rates in proportion of their abuse of the network. A malicious flow continuously hogs more than its fair share of link bandwidth. So LUBA assigns the drop probability to a malicious flow so that it does not get more than its fair share of network. LubaInterval, B, is the byte-count of total packets received by the congested router during an interval to measure whether a flow is hogging more than its fair share. Overload-factor (U) is computed by B bytes arriving at the router. If the overload-factor U is below target link utilization router is non-congested and packets are not marked or dropped otherwise all arriving packets are monitored while assigning a flowId to each ingress flow at the router. A history table is maintained to monitor flows which take more than their fair share of bandwidth in a lubaInterval. It disciplines malicious flows in proportion to their excess inflow. It offers high throughput and avoids global synchronization of responsive flows. LUBA works well in different network conditions and the complexity of the algorithm does not increase even when there is large number of non-responsive flows

*3) OTHERS*

*a) SFB:* It [7] is a FIFO queueing algorithm that identifies and rate-limits non-responsive flows based on accounting mechanisms. The accounting bins are used to keep track of queue occupancy statistics of packets belonging to a particular bin. Each bin keeps a dropping probability $p_m$ which is updated based on bin occupancy. As a packet arrives at the queue, it is hashed into one of the N bins in each of the levels. If the number of packets mapped to a bin goes above a certain threshold, $p_m$ for the bin is increased. If the number of packets drops to zero, $p_m$ is decreased. SFB is highly scalable and enforces fairness using an extremely amount of state and a small amount of buffer space.

*C. Only flow information*

The third category of AQMs uses only the flow information and does not identify the congestion metric to control the congestion.

*a) Stabilised RED:* SRED in [20] pre-emptively discards packets with a load-dependent probability when a buffer in a router is congested. It stabilizes its buffer occupancy at a level independent of the number of the active connections. SRED does this by estimating the number of active connections. It obtains the estimate without collecting or analysing state information. Whenever a packet arrives at the buffer, the arriving packet with randomly chosen packet that recently preceded it into the buffer is compared. The information about the arriving packets is augmented with a "Zombie list". As packets arrive, as long as the list is not full, for every packet the packet flow identifier is added to the list. Once the zombie is full, whenever a packet arrives, it is compared with a randomly chosen zombie in the zombie list. If the arriving packet's flow matches the zombie it is declared "hit". If the two are not of the same flow, it is declared "no hit". The drop probability depends on whether there was a hit or not. This identifies the no. of active flows and finds candidates for misbehaving flow. SRED keeps the buffer occupancy close to a specific target and away from overflow or underflow. In SRED the buffer occupancy is independent of the number of connections while in RED the buffer occupancy increases with the number of connections. The hit mechanism is used to identify misbehaving flows without keeping per-flow state. Stabilised RED overcomes the scalability problem but suffers from low throughput.

*b) GREEN:* This algorithm uses flow parameters and the knowledge of TCP end-host behavior to intelligently mark packets to prevent queue build up, and prevent congestion from occurring. It offers a high utilization and a low packet loss. An improvement of this algorithm is that there are no parameters that need to be tuned to achieve optimal performance in a given scenario. In this algorithm, both the number of flows and the Round Trip Time of each flow are taken into consideration to calculate the congestion-notification probabilities. The marking probability in GREEN is generally different for each flow because it depends on characteristics that are flow specific.

## III. DISCUSSION

In the recent years many AQM mechanisms have been developed which tries to solve the Internet congestion that exists in routers. The various problems like lock-out, global synchronization and fairness are the issues that are considered in these AQMs. To solve these problems, these AQMs used a mixture of concepts. In the previous section, these existing AQM schemes were classified to perform the analysis of AQMs with ease. According to the classification, basically most of the AQMs employed only congestion metric to detect the congestion. However some of the AQMs required additional flow information other than the congestion metric to know the accurate status of the queue. Very few of the AQMs required only the flow information to spot out the congestion. Considering these AQMs relevant to classification, the first category AQMs based only on congestion metric without flow information were more simple and easy to design compared to the second category AQMs based on congestion metric with flow information. However, the second category AQMs also required extra overhead and implementation compared to the first category AQMs. The third category AQMs has a still greater complexity in identifying the flow information for calculating the marking probability. The Table I also projects the AQMs queue occupation status. Most of the AQMs tried to keep the queue size around a target rather than maximizing or minimizing the queue. AQMs that tried to have the queue size around a target performed better than the other AQMs. RED is the first widely employed AQM which detects congestion using only the congestion metric and without flow information. The Table I indicate that irrespective of the congestion indicator additional flow information gives better strength in bring awareness of congestion in routers. Based on RED AQM, many variant AQMs were developed. RED AQM uses multiple parameters that are to be fined tuned. So RED faced this problem of parameter tuning. As a result packet loss and utilization at the link varied with regard to the network load variation. Network load variation also leads to the existence of global synchronization. RED based AQMs like DSRED, MRED, AdaptiveRED tried to remove the problems of RED. DSRED, MRED showed better performance than RED. AdaptiveRED tried to eliminate the problem of parameter tuning by adapting the parameters. Though RED and its variant were simple to handle, the difficulty with it is the parameter tuning problem.

RED based AQMs are vulnerable to unresponsive flows dominating a routers queue. To overcome this problem, FRED was proposed that improved uniformity by constraining all flows to occupy loosely equal shares of the queue's capacity. It removed the problem of unresponsive flows dominating a queue. Though it used the congestion metric, it also had to keep track of the additional flow information to control congestion. This became the major weakness of the FRED. Based on this AQMs were developed to get rid of the overhead. Combination of Flow and congestion metric based AQMs like CHOKe, SFB, SFED, FABA, StoRED were proposed to allocate fair buffer between flows considering the effects of misbehaving or non-responsive flows. CHOKe provides much better fairness than FRED but penalizes high bandwidth flows and does not handle unresponsive flows in case of few packets.

Flow based AQMs with congestion metric are able to discriminate responsive and non-responsive flows. The malicious flows are identified which might cause congestion at the router. Stochastic RED is based on the concept of flow-based AQM and simple, powerful RED algorithm. To avoid

*(IJCSIS) International Journal of Computer Science and Information Security,*
*Vol. 8, No. 1, 2010*maintaining per flow state as in other flow-based AQM, StoRED uses the idea of the time varying hash function to map flows to different counting buns. StoRED is outstanding in disciplining misbehaving flow, making unresponsive flows TCP-friendly, and improving the responsive of Web transfers.

Further these AQMs were classified based on the congestion metric. Most of the AQMs used congestion metric to detect congestion. A variety of congestion indicators like queue length, input rate, packet loss and link utilization were used for congestion detection. RED based AQMs used queue length as congestion indicator. Some of the AQMs tried to prove that Queue status does not give a clear status of the congestion.

REM used both input rate and queue length that illustrated very high utilization but very low throughput compared to Queue based RED. AVQ and YELLOW used only input rate as the congestion indicator to demonstrate that it performed well in terms of link utilisation and packet loss.

BLUE used packet loss and link utilization as congestion indicator to give a very high throughput and, high utilisation with low queue ngth stability. The Table II indicates that the Load- based AQMs perform better with high link utilisation, throughput compared to the Queue-based AQMs. The Table II indicates that irrespective of the additional flow information, Load-based AQMs gives better strength in bringing awareness of congestion in routers.

While comparing the variety of congestion indicators, Queue based AQMs are simple to design except for the parameter tuning problem compared with the other AQMs. Irrespective of the AQMs that depended on flow information Load based AQMs performed better than Queue-based in terms of high throughput and utilization. FABA is a rate based AQM exhibited high throughput compared to FRED and CHOKe by maintaining per active flow state and low implementation complexity. SFB is also a flow-based AQM, an improved version of BLUE. This AQM also shows better fairness compared to BLUE. GREEN, SRED AQMs requires only flow information to sense the congestion in routers. GREEN demonstrates very low utilization and high loss compared to the other AQMs. This study indicates the most of the AQMs used queue length or input rate as their congestion indicators. While using the flow information, the AQMs used either queue length or input rate and not both. AQMs can be designed that uses both queue length and input rate as congestion metric with flow information. . So an AQM can be designed that has advantages of Queue-based AQMs, Load-based AQMs and AQMs with flow information.

TABLE I Comparison of AQM schemes based on Classification

| Classification | | AQM Schemes | Queue Occupation | | | Handling Traffic | | | |
|---|---|---|---|---|---|---|---|---|---|
| | | | Max. the queue Occupation | Min. the queue Occupation | Keeping Queue around a target | Adaptive | Non-Adaptive | | |
| | | | | | | | Robust | Fragile | Nonresponsive |
| Congestion Metric Without Flow Information | Queue-based | RED | × | × | ✓ | ✓ | × | × | × |
| | | ARED, LRED | × | × | ✓ | ✓ | × | × | × |
| | Queue and Load-based | REM | × | ✓ | × | ✓ | × | × | × |
| | Load-Based | YELLOW | × | × | ✓ | ✓ | ✓ | ✓ | ✓ |
| | | AVQ | × | × | ✓ | ✓ | × | × | × |
| | Others | BLUE | ✓ | × | × | ✓ | × | × | × |
| Congestion Metric With Flow Information | Queue-based | FRED | × | × | ✓ | ✓ | ✓ | ✓ | ✓ |
| | | CHOKe | × | × | ✓ | ✓ | ✓ | ✓ | ✓ |
| | | StoRED | × | × | ✓ | ✓ | ✓ | ✓ | ✓ |
| | Others | SFB | ✓ | × | × | ✓ | ✓ | ✓ | ✓ |
| | Load-Based | FABA | × | × | ✓ | ✓ | ✓ | ✓ | ✓ |
| Only Flow Information | | GREEN | × | ✓ | × | ✓ | × | ✓ | ✓ |

338　　http://sites.google.com/site/ijcsis/
ISSN 1947-5500

TABLE II Comparison of AQM schemes based on Performance Metrics

| AQM | Link Utilisation | Throughput | Loss Rate | Queue stability | Fairness | Complexity, Computation |
|---|---|---|---|---|---|---|
| RED | High | Low | High | Moderate | Low | High |
| ARED,LRED | High | Moderate | Moderate | High | Low | High |
| REM | High | Very Low | Low | Very Low | Low | High |
| YELLOW | Very High | Low | Very Low | High | Low | High |
| AVQ | Very High | High | Low | Moderate | Low | High |
| BLUE | High | Very High | Moderate | Low | Low | Moderate |
| FRED | High | High | Low | Moderate | High | Very High |
| CHOKe | Moderate | Moderate | Moderate | Moderate | Moderate | Moderate |
| StoRED | High | High | Low | Moderate | Very High | High |
| SFB | High | Moderate | Moderate | Moderate | Moderate | High |
| FABA | Very High | Very High | Low | High | Very High | Very High |
| GREEN | Very Low | Moderate | High | High | Low | Very High |

IV. CONCLUSION

In this paper, the AQM algorithms are classified based on congestion metrics and the flow information. Most of the AQMs only require congestion indicators while some of them require both congestion indicator and flow information. Very few require only flow information for detecting congestion. These AQMs are compared based on the various performance metrics. This paper tries to project the desirable quality and shortcoming that exists in each of the algorithm in terms of their performance.

It also summarizes the functioning of each algorithm. The simplicity of Queue based algorithms can be improved by using the additional flow information without much existence of the overhead. Better AQM algorithms can be proposed that uses the better features of these algorithms while removing the poor features of it to give the best AQM algorithm.

AUTHORS PROFILE

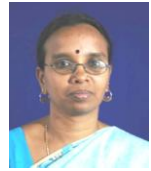

**K.Chitra** received her B.Sc (C.Sc) from Women Christian College, Chennai and M.Sc from Avinashilingam University for Women, Coimbatore in 1991 and 1993 respectively. And, she received her M.Phil degree in Computer Science from Bharathiar University, Coimbatore in 2005. She is pursuing her PhD at Avinashilingam University for Women. She is currently working as a Lecturer in the Department of Computer Science, D.J.Academy for Managerial Excellence, Coimbatore. She has 12 years of teaching experience. Her research interests are Congestion Control in Networks and Network Security.

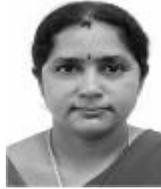

**Dr. Padmavathi Ganapathi** is the Professor and Head of the Department of Computer Science, Avinashilingam University for Women, Coimbatore. She has 21 years of teaching experience and one year Industrial experience. Her areas of interest include Network security and Cryptography and real time communication. She has more than 80 publications at national and International level. She is a life member of many professional organizations like CSI, ISTE, AACE, WSEAS, ISCA, and UWA. She is currently the Principal Investigator of 5 major projects under UGC and DRDO